\documentclass[pra,twocolumn,showpacs,preprintnumbers,amsmath,amssymb]{revtex4}
\usepackage{graphicx}% Include figure files
\usepackage{dcolumn}% Align table columns on decimal point
\usepackage{bm}% bold math
\usepackage{color}

\begin{document}

\preprint{}

\title{Comment on "Information flow of quantum states interacting with closed timelike curves"}% Force line breaks with \\

\author{Waldemar K{\l}obus}

\affiliation{Faculty of Physics, Adam Mickiewicz University, ul. Umultowska 85, 61-614 Pozna\'{n}, Poland}

\author{Andrzej Grudka}

\affiliation{Faculty of Physics, Adam Mickiewicz University, ul. Umultowska 85, 61-614 Pozna\'{n}, Poland}

\author{Antoni W\'{o}jcik}

\affiliation{Faculty of Physics, Adam Mickiewicz University, ul. Umultowska 85, 61-614 Pozna\'{n}, Poland}%Lines break automatically or can be forced with \\

\date{\today}% It is always \today, today,

\begin{abstract}
We show that recent results on the interaction of causality-respecting particles with particles on closed timelike curves derived in [Phys. Rev. A {\bf 82}, 062330 (2010)] depend on ambiguous assumption about the form of the state which is inputted into the proposed equivalent circuit. Choosing different form of this state leads to opposite conclusion on the power of closed timelike curves.
\end{abstract}

\maketitle

In an interesting paper Ralph and Myers considered information flow through closed timelike curves \cite{PhysRevA.82.062330}. To achieve the aim they modeled closed timelike curves by the equivalent circuit which consists of an infinite sequence of identical unitary operations with two inputs and two outputs. One output of each unitary operation in the sequence is identified with one input of the next unitary operation in the sequence. Based on this model they concluded that closed timelike curves did not destroy classical correlations and could distinguish nonorthogonal quantum states. In this Comment we show that their conclusion is based on ambiguous assumption about the form of the state half of which one inputs into the equivalent circuit. Choosing different form of this state leads to conclusion that closed timelike curves destroy classical correlations and cannot help in distinguishing nonorthogonal states.

Let us first describe how different states can be represented in the equivalent circuit formalism.
Suppose that we input half of the pair of particles in a pure state $|\phi\rangle_{AB}$ into one arm of the wormhole represented by unitary operator $U$ with the lower input and the upper output  identified (see Fig. 3 (a) in \cite{PhysRevA.82.062330}). In the equivalent circuit model it corresponds to preparation of $n$ pairs of particles in a pure initial state $|\phi\rangle_{AB}^{\otimes n}$ and inputting $n$ B-particles into the equivalent circuit (see Fig. 3 (b) in \cite{PhysRevA.82.062330}). After the action of the circuit one traces out all particles except the two ones which correspond to a single unitary operator $U$. More precisely, one leaves the $A$-particle corresponding to open incoming line and one leaves the $B$ particle corresponding to open outgoing line of the same unitary operator $U$.

Suppose now that we input into one arm of the wormhole half of the pair of particles in a mixed state $\rho=\sum_i p_i |\phi_i\rangle\langle\phi_i|_{AB}$. The authors of \cite{PhysRevA.82.062330} assume that it corresponds to preparation of $n$ pairs of particles in an initial state
\begin{eqnarray}
\rho=\sum_i p_i |\phi_i\rangle\langle\phi_i|_{AB}^{\otimes n}.
\label{eq:1}
\end{eqnarray}
However one may as well assume that it corresponds to preparation of $n$ particles in an initial state
\begin{eqnarray}
\rho=(\sum_i p_i |\phi_i\rangle\langle\phi_i|_{AB})^{\otimes n}.
\label{eq:2}
\end{eqnarray}
In both cases one inputs $n$ $B$-particles into the equivalent circuit and then proceeds in the way described in \cite{PhysRevA.82.062330}.
Because both states lead to identical reduced density matrices of a single pair of particles one cannot unambiguously decide which is valid. In particular the states are identical if they represent pure states (if some $p_i$ goes to $1$ then both states go to $|\phi_i\rangle\langle\phi_i|_{AB}^{\otimes n}$).

One could decide which state is valid if only one knew whether the dynamics of states interacting with closed timelike curves depends only on a reduced density matrix of the state, or -- in the case of proper mixtures -- on a particular ensemble realizing the reduced density matrix. If the dynamics depends only on a reduced density matrix of the state then the state of Eq. \ref{eq:2} is valid. On the other hand, if the dynamics depends on a particular ensemble realizing the reduced density matrix then the state of Eq. \ref{eq:1} is valid. However, at the moment it is not known what the dynamics of states interacting with closed timelike curves depends on.

Before we show that different forms of initial state lead to different results we recall for further convenience the argument given in \cite{PhysRevA.82.062330} in the case when one inputs half of the maximally entangled pair of particles into one arm of the wormhole represented by unitary operator $U=I$ ($I$ denotes identity operator) with the lower input and the upper output  identified (see Fig. 3 (a) in \cite{PhysRevA.82.062330}). It corresponds to preparation of $n$ maximally entangled pairs of particles, i.e. a state $|\phi^+\rangle_{AB}^{\otimes n}$, where $|\phi^+\rangle_{AB}=1/\sqrt{2}(|00\rangle_{AB}+|11\rangle_{AB})$ and inputting $n$ $B$-particles into the equivalent circuit (see Fig. 3 (b) in \cite{PhysRevA.82.062330}). We trace out all particles except two particles which correspond to a single unitary operator $U$. As the two particles originate from two different entangled pairs the equivalent circuit decorrelates particles.

Let us now consider that we input half of the maximally classically correlated pair of particles into one arm of the wormhole. The authors of \cite{PhysRevA.82.062330} assume that it corresponds to preparation of a state
\begin{eqnarray}
\frac{1}{2}|00\rangle\langle00|_{AB}^{\otimes n}+\frac{1}{2}|11\rangle\langle11|_{AB}^{\otimes n}
\label{eq:3}
\end{eqnarray}
and inputting $n$ $B$-particles into the equivalent circuit. In consequence they obtain that classical   correlations are not destroyed by the wormhole.
However we may assume that instead of a state of Eq. \ref{eq:3} we prepare a state
\begin{eqnarray}
\left(\frac{1}{2}|00\rangle\langle00|_{AB}+\frac{1}{2}|11\rangle\langle11|_{AB}\right)^{\otimes n}
\label{eq:4}
\end{eqnarray}
and input $n$ $B$-particles into the equivalent circuit.  We trace out all particles except two particles which correspond to a single unitary operator $U$. We can now argue as the authors of \cite{PhysRevA.82.062330} did in the case of maximally entangled pairs  that the two particles originate from two different classically correlated pairs and the equivalent circuit decorrelates particles. Hence, contrary to the conclusion in \cite{PhysRevA.82.062330} we obtain that classical correlations are destroyed by the wormhole.

It is also instructive to analyze the following thought experiment \footnote{We thank our Referee for suggesting this example.}. We prepare the maximally entangled pair of particles and input the $B$-particle into one arm of the wormhole. However, before we input the $B$-particle into one arm of the wormhole we measure the $A$-particle in a basis $\{|0\rangle_A ,|1\rangle_A\}$. In the equivalent circuit model we can describe this process as follows. We prepare $n$ maximally entangled pairs of particles, i.e. a state $|\phi^+\rangle_{AB}^{\otimes n}$. Because in the equivalent circuit model one traces out all particles except two particles which correspond to a single unitary operator $U$ one can argue that one can measure only one $A$-particle. Hence after the measurement we obtain a state
\begin{eqnarray}
& |\phi^+\rangle \langle\phi^+|_{AB}^{\otimes k}(\frac{1}{2}|00\rangle\langle00|_{AB}+\nonumber\\
& +\frac{1}{2}|11\rangle\langle11|_{AB})|\phi^+\rangle \langle\phi^+|_{AB}^{\otimes n-k-1},
\label{eq:5}
\end{eqnarray}
where we {\it do not} trace out the $k+1$-th $A$-particle and the $k$-th $B$-particle. Now we input $B$-particles into the equivalent circuit. Although the state of Eq. \ref{eq:5} is different than the state of Eq. \ref{eq:4} both states lead to identical results, i.e., we obtain that classical correlations are destroyed by the wormhole. On the other hand in order to obtain the state of Eq. \ref{eq:3} one should measure all $A$-particles and postselect a state which corresponds to results of measurements $|0\rangle_A^{\otimes n}$ or $|1\rangle_A^{\otimes n}$.

Finally let us consider the case of distinguishing nonorthogonal states $|0\rangle$ and $|-\rangle=\frac{1}{\sqrt{2}}(|0\rangle -|1\rangle)$. We prepare a state $\frac{1}{2}|00\rangle\langle00|_{AB}+\frac{1}{2}|1-\rangle\langle1-|_{AB}$ and input the $B$-particle into one arm of the wormhole represented by unitary operator $U=\textrm{\textsc{ch}}$ ($\textrm{\textsc{ch}}$ denotes controlled Hadamard operator with control on the lower particle) with the lower input and the upper output identified. The authors of \cite{PhysRevA.82.062330} assume that it corresponds to preparation of a state
\begin{eqnarray}
\frac{1}{2}|00\rangle\langle00|_{AB}^{\otimes n}+\frac{1}{2}|1-\rangle\langle1-|_{AB}^{\otimes n}
\label{eq:6}
\end{eqnarray}
and inputting $n$ $B$-particles into the equivalent circuit. In consequence they obtain the result of Brun et al. \cite{PhysRevLett.102.210402}, i.e. closed timelike curves can distinguish nonorthogonal states.
However, if instead we assume that it corresponds to preparation of a state
\begin{eqnarray}
\left(\frac{1}{2}|00\rangle\langle00|_{AB}+\frac{1}{2}|1-\rangle\langle1-|_{AB}\right)^{\otimes n}
\label{eq:7}
\end{eqnarray}
and inputting $n$ $B$-particles into the equivalent circuit then we obtain the result of Bennett et al. \cite{PhysRevLett.103.170502}, i.e. closed timelike curves cannot help in distinguishing nonorthogonal states.
It is not surprising that in the first case we succeeded in distinguishing nonorthogonal states because the reduced state of $B$-particles is a mixture of two pure states $|0\rangle_{B}^{\otimes n}$ and $|-\rangle_{B}^{\otimes n}$, which are almost orthogonal for large $n$. On the other hand, in the second case the reduced state of $B$-particles is a mixture of $2^n$ nonorthogonal pure states.

A.G. was partially supported by the Polish Ministry of Science and Higher Education Grant No. N N202 231937.

\end{document}